\documentclass[CEJP,PDF]{cej} 
\usepackage{layout}
\usepackage{amsmath}
\usepackage{textcomp}
\usepackage{hyperref}

\title{Is gravitational mass of a composite quantum body equivalent to its energy?}

\articletype{Research Article}

\author{Andrei~G.~Lebed\inst{1}$^,$\inst{2}\email{lebed@physics.arizona.edu}}

\institute{
     \inst{1} Department of Physics, University of Arizona,\\
     1118 E. 4-th Street, Tucson, AZ 85721 , USA
     \inst{2} L.D. Landau Institute for Theoretical Physics, RAS,\\
     2 Kosygina Street, 117334 Moscow, Russia}

\abstract{We define passive gravitational mass operator of a hydrogen atom in the
post-Newtonian approximation of general relativity and show that it does
not commute with energy operator, taken in the absence of gravitational field.
Nevertheless, the equivalence between the expectation values of passive gravitational
mass and energy is shown to survive for stationary quantum states. Inequivalence between
passive gravitational mass and energy at a macroscopic level results in time dependent
oscillations of the expectation values of passive gravitational mass for superpositions
of stationary quantum states, where the equivalence restores after averaging over time.
Inequivalence between gravitational mass and energy at a microscopic level reveals
itself as unusual electromagnetic radiation, emitted by the atoms, supported and moved
in the Earth gravitational field with constant velocity using spacecraft or satellite,
which can be experimentally measured.
}


\begin{document}
\maketitle


\section{Introduction}

One of the major problems in physics is known to be a creation of the so-called
"Theory of Everything" - unification of all fundamental forces in nature, including
electro-weak, strong, and gravitational ones. In this context, the most difficult
step is a creation of quantum gravitational theory. This even may not be possible
in the feasible future, since the fundamentals of quantum mechanics are very different
from that of general relativity. In this situation, it is important to find a way
to combine quantum mechanics with some non-trivial approximation of the general relativity.
This allows to introduce some novel physical ideas and phenomena, which can be experimentally
studied. So far, to the best of our knowledge, only trivial quantum mechanical variant of
the Newton approximation of general relativity has been experimentally tested in the famous
COW [1] and ILL [2] experiments. On the other hand, such important and non-trivial
quantum phenomena in general relativity as the Hawking radiation [3] and the Unruh effect
[4] are still far from their direct and firm experimental confirmations.

It is known that gravitational mass of a composite classical body
in general relativity is not a trivial notion and is a subject of
several paradoxes. One of them is related to application of the
so-called Tolman formula [5] to a free photon, which formally
results in a doubling of photon active gravitational mass [5,6].
The solution of this paradox is due to an account of stress in the
walls of a container [6], containing photon, which compensates the
above mentioned increase of photon mass. More precisely, it is
shown [6] that averaged over time active gravitational mass of a
photon in a container with mirror walls is $E/c^2$ , where $E$ is
photon energy. Importance of the classical virial theorem for the
equivalence between averaged over time gravitational mass and
energy of different composite classical bodies is stressed in
[7,8]. In particular, it is shown that electrostatic binding
energy, $U$, contributes to passive and active gravitational
masses as $2U/c^2$, whereas kinetic energy, $K$, contribution to
the gravitational masses is $3K/c^2$ [7,8]. Application of the
classical virial theorem to electrostatitally bound two bodies,
which claims that  averaged over time potential energy, $<U>_t$,
equals to -$2<K>_t$, results in the equivalence between averaged
over time gravitational mass and energy,
\begin{equation}
<m>_t= m_1 + m_2 + E/c^2,
\end{equation}
where $m_1$ and $m_2$ are bare masses of the above mentioned bodies,
$E=K+U$.

\section{Goal}

The main goal of our paper is to consider a quantum mechanical
problem about interaction of a composite body (e.g., a hydrogen
atom) with an external gravitational field (e.g., the Earth). Our
first result is that the equivalence between passive gravitational
mass and energy may survive at a macroscopic level. In particular,
we show that the quantum virial theorem [9] results in equality
between the expectation value of passive gravitational mass and
energy, divided into $c^2$, for stationary quantum states. Our
second result is a breakdown of the equivalence between passive
gravitational mass and energy at a microscopic level. We define
passive gravitational mass operator of a hydrogen atom in the
post-Newtonian approximation of general relativity and show that
it does not commute with energy operator, taken in the absence of
gravitational field. Therefore, an atom with definite energy in
the absence of gravitational field, $E$, is not characterized by
definite passive gravitational mass in an external gravitational
field. Passive gravitational mass is shown to be quantized and can
significantly differ from $E/c^2$. Our third result is that we
discuss how the above mentioned inequivalence can be
experimentally observed. In particular, we suggest experimental
investigation of electromagnetic radiation, emitted by macroscopic
ensemble of the atoms [10,11], supported and moved with constant
velocity in the Earth gravitational field by using spacecraft or
satellite. Our fourth result is that the equivalence between the
expectation values of passive gravitational mass and energy is
broken at macroscopic level for superpositions of stationary
quantum states and restores only after averaging over time.

\section{Gravitational mass in classical physics}

Here, we derive the Lagrangian and Hamiltonian of a hydrogen atom
in the Earth gravitational field, taking into account couplings of
kinetic and potential Coulomb energies of an electron with a weak
centrosymmetric gravitational field. Note that we keep only terms
of the order of $1/c^2$ and disregard magnetic force, radiation of
both electromagnetic and gravitational waves as well as all tidal
and spin dependent effects. Let us write the interval in the Earth
gravitational field using the so-called weak field approximation
[12]:
\begin{equation}
d s^2 = -\biggl(1 + 2 \frac{\phi}{c^2} \biggl)(cdt)^2 + \biggl(1 - 2
\frac{\phi}{c^2} \biggl) (dx^2 +dy^2+dz^2 ), \ \phi = - \frac{GM}{R}
,
\end{equation}
where $G$ is the gravitational constant, $c$ is the velocity of
light, $M$ is the Earth mass, $R$ is a distance between a center of
the Earth and a center of mass of a hydrogen atom (i.e., proton). We
pay attention that to calculate the Lagrangian (and later - the
Hamiltonian) in a linear with respect to a small parameter $\phi(R)
/ c^2$ approximation, we do not need to keep the terms of the order
of $[\phi(R)/c^2]^2$ in metric (2), in contrast to the perihelion
orbit procession calculations [12] .

Then, in the local proper spacetime coordinates,
\begin{equation}
x'=\biggl(1-\frac{\phi}{c^2} \biggl) x, \ y'=
\biggl(1-\frac{\phi}{c^2} \biggl) y, z'=\biggl(1-\frac{\phi}{c^2}
\biggl) z , \ t'= \biggl(1+\frac{\phi}{c^2} \biggl) t,
\end{equation}
the classical Lagrangian and action of a hydrogen atom have the
following standard forms:
\begin{equation}
L' = -m_p c^2 -m_e c^2 + \frac{1}{2} m_e ({\bf v'})^2 +
\frac{e^2}{r'} \ , \ \ \ S' = \int L' dt' ,
\end{equation}
where $m_p$ and $m_e$ are the bare proton and electron masses,
respectively; $e$ and ${\bf v'}$ are the electron charge and
velocity, respectively; $r'$ is a distance between electron and
proton. [Note that, due to inequality $m_p \gg m_e$, we disregard
kinetic energy of a proton in the Lagrangian (4) and consider its
position as a position of center of mass of a hydrogen atom]. It
is possible to show that the Lagrangian (4) can be rewritten in
coordinates $(x,y,z,t)$ as
\begin{equation}
L = -m_p c^2 -m_e c^2 +  \frac{1}{2}m_e{\bf v}^2+\frac{e^2}{r} -
m_e \phi - \biggl( 3m_e\frac{{\bf v}^2}{2}-2\frac{e^2}{r} \biggl)
\frac{\phi}{c^2} .
\end{equation}

Let us calculate the Hamiltonian, corresponding to the Lagrangian
(5), by means of a standard procedure, $H({\bf p},{\bf r})={\bf
p}{\bf v}-L({\bf v},{\bf r})$, where ${\bf p}= \partial L({\bf
v},{\bf r})/\partial {\bf v}$. As a result, we obtain:
\begin{equation}
H = m_p c^2 + m_e c^2 + \frac{{\bf p}^2}{2m_e}-\frac{e^2}{r} + m_p
\phi + m_e \phi + \biggl( 3 \frac{{\bf p}^2}{2 m_e}
-2\frac{e^2}{r} \biggl) \frac{\phi}{c^2},
\end{equation}
where canonical momentum in a gravitational field is ${\bf p}=m_e
{\bf v}(1-3\phi/c^2)$. [Note that, in the paper, we disregard all
tidal effects (i.e., we do not differentiate gravitational
potential with respect to electron coordinates, ${\bf r}$ and
${\bf r'}$, corresponding to a position of an electron in the
center of mass coordinate system). It is possible to show that
this means that we consider the atom as a point-like body and
disregard all effects of the order of $|\phi/c^2|(r_B/R) \sim
10^{-26}$, where $r_B$ is the Bohr radius (i.e., a typical size of
the atom).] From the Hamiltonian (6), averaged over time hydrogen
atom passive gravitational mass, $<m^g>_t$, defined as its weight
in a weak centrosymmetric gravitational field (2), can be
expressed as
\begin{equation}
<m^g>_t = m_p + m_e  + \biggl<  \frac{{\bf p}^2}{2 m_e}-
\frac{e^2}{r}\biggl>_t \frac{1}{c^2} +\biggl< 2 \frac{{\bf p}^2}{2
m_e}-\frac{e^2}{r}\biggl>_t \frac{1}{c^2} =  m_p + m_e +
\frac{E}{c^2} \ ,
\end{equation}
where $E= {\bf p}^2/2 m_e - e^2/r$ is an electron energy. We pay
attention that averaged over time third term in Eq.(7) is equal to
zero due to the classical virial theorem. Thus, we conclude that in
classical physics averaged over time passive gravitational mass of a
composite body is equivalent to its energy, taken in the absence of
gravitational field [7,8].

\section{Gravitational mass in quantum physics}

Now, we proceed to the original part of our paper. The Hamiltonian
(6) can be quantized by substituting a momentum operator,
$\hat{\bf p} = - i \hbar
\partial /\partial {\bf r}$, instead of canonical momentum, ${\bf
p}$. It is convenient to write the quantized Hamiltonian in the
following form:
\begin{equation}
\hat H = m_p c^2 + m_e c^2 + \frac{\hat {\bf
p}^2}{2m_e}-\frac{e^2}{r} + m_p \phi + \hat m^g_e \phi \ ,
\end{equation}
where we introduce passive gravitational mass operator of an
electron to be proportional to its weight operator in a weak
centrosymmetric gravitational field (2),
\begin{equation}
\hat m^g_e  = m_e + \biggl(\frac{\hat {\bf p}^2}{2m_e}
-\frac{e^2}{r}\biggl)\frac{1}{c^2} + \biggl(2 \frac{\hat {\bf
p}^2}{2m_e}-\frac{e^2}{r} \biggl) \frac{1}{c^2} \ .
\end{equation}
Note that the first term in Eq.(9) corresponds to the bare
electron mass, $m_e$, the second term corresponds to the expected
electron energy contribution to the mass operator, whereas the
third nontrivial term is the virial contribution to the mass
operator. It is important that the operator (9) does not commute
with electron energy operator, taken in the absence of the field.
It is possible to show that Eqs.(8),(9) can be also obtained
directly from the Dirac equation in a curved spacetime,
corresponding to a weak gravitational field (2). For example, the
Hamiltonian (8),(9) can be obtained [10,11] from the Hamiltonian
(3.24) of Ref.[13] (where different physical problem is
considered) by omitting all tidal terms.

\subsection{Equivalence of the expectations values}

Below, we discuss some consequences of Eq.(9). Suppose that we have
a macroscopic ensemble of hydrogen atoms with each of them being in
a ground state with energy $E_1$. Then, as follows from Eq.(9), the
expectation value of the gravitational mass operator per one electron is
\begin{equation}
<\hat m^g_e> = m_e + \frac{ E_1}{c^2}  + \biggl< 2 \frac{\hat {\bf
p}^2}{2m_e}-\frac{e^2}{r} \biggl> \frac{1}{c^2} = m_e +
\frac{E_1}{c^2}  ,
\end{equation}
where the third term in Eq.(10) is zero in accordance with the
quantum virial theorem [9]. Therefore, we conclude that the
equivalence between passive gravitational mass and energy in the
absence of gravitational field survives at a macroscopic level for
stationary quantum states.

\subsection{Breakdown of the equivalence at a microscopic level}

Let us discuss how Eqs.(8),(9) break the equivalence between passive
gravitational mass and energy at a microscopic level. First of all,
we recall that the mass operator (9) does not commute with electron
energy operator, taken in the absence of gravitational field. This
means that, if we create a quantum state of a hydrogen atom with
definite energy, it will not be characterized by definite passive
gravitational mass. In other words, a measurement of the mass in
such quantum state may give different values, which, as shown, are
quantized. Here, we illustrate the above mentioned inequivalence
using the following thought experiment. Suppose that, at $t=0$, we
create a ground state wave function of a hydrogen atom,
corresponding to the absence of gravitational field,
\begin{equation}
\Psi_1(r,t) = \Psi_1(r) \exp(-iE_1t/\hbar) \ .
\end{equation}
In a weak centrosymmetric gravitational field (2), wave function (11)
is not anymore a ground state of the Hamiltonian (8),(9), where we
treat gravitational field as a small perturbation in an inertial
system [7,8,10-13]. It is important that for an inertial observer, in
accordance with Eq.(3), a general solution of the Schr\"{o}dinger
equation, corresponding to the Hamiltonian (8),(9), can be written
as
\begin{equation}
\Psi(r,t) = (1-\phi/c^2)^{3/2} \sum^{\infty}_{n = 1} a_n \Psi_n [(1-
\phi/c^2)r]
\ \exp[-i m_e c^2 (1+\phi/c^2) t/\hbar]
\ \exp[-i E_n(1+\phi/c^2) t/\hbar] \ .
\end{equation}
We pay attention that wave function (12) is a series of
eigenfunctions of passive gravitational mass operator (9), if we
take into account only linear terms with respect to the parameter
$\phi/c^2$. Here, factor $1-\phi/c^2$ is due to a curvature of
space, whereas the term $E_n(1+\phi/c^2)$ represents the famous
red shift which is an effect of the gravitational field on time.
$\Psi_n(r)$ is a normalized wave function of an electron in a
hydrogen atom in the absence of gravitational field, corresponding
to energy $E_n$. [Note that, due to symmetry of our problem, an
electron from $1S$ ground state of a hydrogen atom can be excited
only into $nS$ excited states. We also pay attention that the wave
function (12) contains a normalization factor
$(1-\phi/c^2)^{3/2}$.]

In accordance with the basic principles of quantum mechanics,
probability that, at $t>0$, an electron occupies excited state with
energy $m_e c^2(1+\phi/c^2) + E_n(1+\phi/c^2)$ is
\begin{equation}
P_n = |a_n|^2, \ a_n = \int \Psi^*_1(r) \Psi_n [(1-\phi/c^2)r] d^3
{\bf r}= - ( \phi/c^2) \int \Psi^*_1(r) r \Psi'_n(r) d^3 {\bf r}.
\end{equation}
Note that it is possible to demonstrate that for $a_1$ in Eq.(13) a
linear term with respect to gravitational potential, $\phi$, is
zero, which is a consequence of the quantum virial theorem. Taking
into account that the Hamiltonian is a Hermitian operator, it is
possible to show that for $n \neq 1$:
\begin{equation}
\int \Psi^*_1(r) r \Psi'_n(r) d^3 {\bf r} = V_{n,1}/ (\hbar
\omega_{n,1}), \ \hbar \omega_{n,1} = E_n-E_1 , \ n \neq 1 ,
\end{equation}
where $V_{n,1}$ is a matrix element of the virial operator,
\begin{equation}
V_{n,1}= \int \Psi^*_1(r) \hat V({\bf r}) \Psi_n(r) d^3 {\bf r} , \
\ \hat V({\bf r}) = 2 \frac{\hat {\bf p}^2}{2 m_e} - \frac{e^2}{r}.
\end{equation}
It is important that, since the virial operator (15) does not
commute with the Hamiltonian, taken in the absence of gravitational
field, the probabilities (13)-(15) are not equal to zero for $n \neq
1$.

Let us discuss Eqs.(12)-(15). We pay attention that they directly
demonstrate that there is a finite probability,
\begin{equation}
P_n = |a_n|^2 = \Big( \frac{\phi}{c^2} \Big)^2 \ \Big(
\frac{V_{n,1}}{E_n-E_1} \Big)^2 \ , \ n \neq 1,
\end{equation}
that, at $t>0$, an electron occupies n-th ($n \neq 1$) energy level,
which breaks the expected Einstein equation, $m^g_e=m_e + E_1/c^2$.
In fact, this means that measurement of passive gravitational mass
(i.e., weight in the gravitational field (2)) in a quantum state with
a definite energy (11) gives the following quantized values:
\begin{equation}
m^g_e (n) = m_e + E_n/c^2  \ ,
\end{equation}
corresponding to the probabilities (16). [Note that, as follows
from quantum mechanics, we have to calculate wave function (12) in a
linear approximation with respect to the parameter $\phi/c^2$ to
obtain probabilities (16), which are proportional to
$(\phi/c^2)^2$. A simple analysis shows that an account in Eq.(12)
terms of the order of $(\phi/c^2)^2$ would change electron passive
gravitational mass of the order of  $(\phi / c^2) m_e \sim 10^{-9}
m_e$, which is much smaller than the distance between the quantized
values (17), $\delta m^g_e \sim \alpha^2 m_e \sim 10^{-4} m_e$,
where $\alpha$ is the fine structure constant.] We also point out
that, although the probabilities (16) are quadratic with respect to
gravitational potential and, thus, small, the changes of the passive
gravitational mass (17) are large and of the order of $\alpha^2
m_e$. We also pay attention that small values of probabilities (16),
$P_n \sim 10^{-18}$, do not contradict the existing E\"{o}tv\"{o}s type
measurements [12], which have confirmed the equivalence principle
with the accuracy of the order of $10^{-12}-10^{-13}$. For our case,
it is crucial that the excited levels of a hydrogen atom
spontaneously decay with time, therefore, one can detect the
quantization law (17) by measuring electromagnetic radiation,
emitted by a macroscopic ensemble of hydrogen atoms. The above
mentioned optical method is much more sensitive than the E\"{o}tv\"{o}s type
measurements and we, therefore, hope that it allows to detect the
breakdown of  the equivalence between energy and passive
gravitational mass, revealed in the paper.

\section{Suggested experiment}

\subsection{Lagrangian and Hamiltonian}

Now, let us consider the Lagrangian of a three body system: a
hydrogen atom and the Earth in an inertial coordinate system,
related to a center of mass of a hydrogen atom, where the Earth
moves from the atom with small and constant velocity $u \ll c$. In
this case, we can make use of the results of Ref.\cite{Nordtvedt},
where the corresponding many-body Lagrangian is calculated as a
sum:
\begin{equation}
L = L_{kin} + L_{em} + L_{G} + L_{e,G},
\end{equation}
where $L_{kin}$, $L_{em}$, $L_G$, and $L_{e,G}$ are kinetic,
electromagnetic, gravitational and electric-gravitational parts of
the Lagrangian, respectively. In our approximation, where we
disregard in the Lagrangian and Hamiltonian all terms of the order
of $(v/c)^4$ and $(\phi/c^2)^2$ as well as keep only classical
kinetic and Coulomb potential energies couplings with external
gravitational field, different contributions to the Lagrangian
(18) can be rewritten from Ref. \cite{Nordtvedt} in the following
simplified way:

\begin{equation}
L_{kin} + L_{em} = - Mc^2 - m_pc^2 - m_ec^2 + M\frac{u^2}{2} +
m_e\frac{{\bf v}^2}{2} + \frac{e^2}{r} \ ,
\end{equation}

\begin{equation}
L_{G} = G \frac{m_p M}{(R+ut)} + G \frac{m_e M}{(R+ut)} \biggl\{1
- \frac{1}{2} [{\bf v} \cdot {\bf u} + ({\bf v} \cdot \hat{\bf
R})({\bf u} \cdot \hat{\bf R})]/c^2 \biggl\} + \frac{3}{2} G
\frac{m_e M}{(R+ut)} \frac{{\bf v^2}}{c^2} \ ,
\end{equation}

\begin{equation}
L_{e,G} = - 2 G \frac{M}{(R+ut)c^2} \frac{e^2}{r} \ ,
\end{equation}
where $\hat{\bf R}$ is a unit vector directed along ${\bf R}$ and
where we use the inequality $m_e \ll m_p$.

Then, accepting and using the following condition,
\begin{equation}
u \ll v \sim \alpha c \ ,
\end{equation}
(where $\alpha c$ is a typical electron velocity in a hydrogen
atom) and keeping only relevant terms in the Lagrangian, we can
write:
\begin{equation}
L = m_e \frac{v^2}{2} + \frac{e^2}{r} - \frac{\phi(R+ut)}{c^2}
\biggl[ m_e + 3 m_e \frac{{\bf v}^2}{2} - 2 \frac{e^2}{r} \biggl]
\ .
\end{equation}
[Note that taking into account that a center of mass of a hydrogen
atom does not exactly coincide with a proton would give in the
Lagrangians (20),(23) some extra terms of the order
$|\phi(R+ut)|m_evu/c^2 \ll |\phi(R+ut)|m_ev^2/c^2$. In addition,
we disregard the difference between electron mass and the
so-called reduced mass, which are almost equal under the condition
$m_e \ll m_p$.] It is easy to show that the corresponding electron
Hamiltonian is
\begin{equation}
H= \frac{{\bf p}^2}{2m_e} -\frac{e^2}{r} + \frac{\phi(R+ut)}{c^2}
\biggl[ m_e + 3 \frac{{\bf p}^2}{2m_e} - 2 \frac{e^2}{r} \biggl] \
,
\end{equation}
which can be quantized as it was done in Sect.4:
\begin{equation}
\hat H= \frac{\hat {\bf p}^2}{2m_e} -\frac{e^2}{r} +
\frac{\phi(R+ut)}{c^2} \biggl[ m_e + 3 \frac{\hat {\bf p}^2}{2m_e}
- 2 \frac{e^2}{r} \biggl] \ .
\end{equation}

\subsection{Photon emission and mass quantization}

Here, we describe a realistic experiment [10,11]. We consider a
hydrogen atom to be in its ground state at $t=0$ and located at
distance $R'$ from a center of the Earth. The wave function of a
ground state, corresponding to Hamiltonian (8),(9), can be written
as
\begin{equation}
\tilde{\Psi}_1(r,t) = (1-\phi'/c^2)^{3/2} \Psi_1[(1-\phi'/c^2)r]
\ \exp[-im_ec^2(1+\phi'/c^2) t /\hbar]
\ \exp[-iE_1(1+\phi'/c^2)t/\hbar] \ ,
\end{equation}
where $\phi'=\phi(R')$. The atom is supported in the Earth
gravitational field and moved from the Earth with constant
velocity, $u \ll \alpha c$ (see Eq.(22)), by spacecraft or
satellite. According to Eq.(25), electron wave function and time
dependent perturbation for the Hamiltonian (8),(9) in this
inertial coordinate system can be expressed as
\begin{equation}
\tilde{\Psi}(r,t) = (1- \phi'/c^2)^{3/2} \sum^{\infty}_{n=1}
\tilde{a}_n(t) \Psi_n[(1-\phi'/c^2)r]
\ \exp[-im_ec^2(1+\phi'/c^2) t
/\hbar]
 \ \exp[-iE_n(1+\phi'/c^2)t/\hbar] ,
\end{equation}
\begin{equation}
\hat U ({\bf r},t) =\frac{\phi(R'+ut)-\phi(R')}{c^2}  \biggl(3
\frac{\hat {\bf p}^2}{2m_e}-2\frac{e^2}{r} \biggl) .
\end{equation}
We pay attention that in a spacecraft (satellite), which moves
with constant velocity, gravitational force, which acts on each
hydrogen atom, is compensated by some non-gravitational forces.
This causes very small changes of a hydrogen atom energy levels
and is not important for our calculations. Therefore, the atoms do
not feel directly gravitational acceleration, ${\bf g}$, but feel,
instead, gravitational potential, $\phi (R'+ut)$, changing with
time due to a spacecraft (satellite) motion in the Earth
gravitational field. Application of the time-dependent quantum
mechanical perturbation theory [9] gives the following solutions
for functions $\tilde a_n(t)$ in Eq.(26):
\begin{equation}
\tilde{a}_n(t)= -\frac{V_{n,1}}{\hbar \omega_{n,1}c^2}
\biggl\{[\phi(R'+ut)-\phi(R')]  \exp(i \omega_{n,1}t) - \frac{u}{i
\omega_{n,1}} \int^t_0 \frac{d \phi(R'+ut)}{dR'} d[\exp(i
\omega_{n,1}t)]\biggl\}, \ n \neq 1 \ ,
\end{equation}
where $V_{n,1}$ and $\omega_{n,1}$ are given by Eqs.(14),(15).
Note that under the suggested experiment the following condition
is obviously fulfilled:
\begin{equation}
u \ll \omega_{n,1} R \sim \alpha c (R/r_B) \sim 10^{13} c,
\end{equation}
therefore, we can keep only the first term in the amplitude (29):
\begin{equation}
\tilde{a}_n(t)= -\frac{V_{n,1}}{\hbar \omega_{n,1}c^2}
[\phi(R'+ut)-\phi(R')]  \exp(i \omega_{n,1}t) \ , \ n \neq 1 .
\end{equation}

As follows from Eq.(31), if excited levels of a hydrogen atom were
strictly stationary, then a probability to find the passive
gravitational mass to be quantized with $n \neq 1$ in Eq.(17)
would be
\begin{equation}
\tilde{P}_n(t)= \biggl( \frac{V_{n,1}}{\hbar \omega_{n,1}}
\biggl)^2 \frac{[\phi(R'+ut)-\phi(R')]^2}{c^4} = \biggl(
\frac{V_{n,1}}{\hbar \omega_{n,1}} \biggl)^2
 \frac{[\phi(R")-\phi(R')]^2}{c^4} \ , n \neq 1,
\end{equation}
where $R"=R'+ut$. In reality, the excited levels spontaneously
decay with time and, therefore, it is possible to observe the
quantization law (17) indirectly by measuring electromagnetic
radiation from a macroscopic ensemble of the atoms. In this case,
Eq.(32) gives a probability that a hydrogen atom emits a photon
with frequency $\omega_{n,1} = (E_n-E_1) / \hbar$ during the time
interval $t$. [We note that dipole matrix elements for $nS
\rightarrow 1S$ quantum transitions are zero. Nevertheless, the
corresponding photons can be emitted due to quadrupole effects.]

It is important that the probabilities (32) depend only on
gravitational potential in initial, $\phi'=\phi(R')$, and final,
$\phi''=\phi(R")=\phi(R'+ut)$, positions of a rocket (satellite).
This allows us to clarify their physical meaning. Let us consider
a general solution of Hamiltonian (8),(9) for a hydrogen atom,
located at a distance $R"$ from the center of the Earth:
\begin{equation}
\tilde{\Psi}(r,t) = (1- \phi"/c^2)^{3/2} \sum^{\infty}_{n=1}
\tilde{a}_n \Psi_n[(1-\phi"/c^2)r] \ \exp[-im_ec^2(1+\phi"/c^2) t
/\hbar]
 \ \exp[-iE_n(1+\phi"/c^2)t/\hbar] .
\end{equation}
[Note that the wave function (33) corresponds to an infinite
number of eighenfunctions of a weight operator at a position
$R"$]. Now, if we ask a question: "What is the probability for
electron with definite energy at position $R'$ [wave function
(26)] to have a quantized weight at position $R"$ with $n \neq 1$
[see Eq.(17)]?", then the answer is given by main principles of
quantum mechanics:
\begin{eqnarray}
&&a_n = \int \Psi^*_1[(1-\phi'/c^2)r] \Psi_n [(1-\phi"/c^2)r] d^3
{\bf r}= - [(\phi"-\phi')/c^2] \int \Psi^*_1(r) r \Psi'_n(r) d^3
{\bf r}\ ,
\nonumber\\
&&P_n = |a_n|^2 = \frac{(\phi"-\phi')^2}{c^4}\biggl|\Psi^*_1(r) r
\Psi'_n(r) d^3 \biggl|^2 \ .
\end{eqnarray}
By using Eq.(14), the probabilities (34) can be written in a
familiar way:
\begin{equation}
P_n= \biggl( \frac{V_{n,1}}{\hbar \omega_{n,1}} \biggl)^2
 \frac{[\phi(R")-\phi(R')]^2}{c^4} \ , n \neq 1 ,
\end{equation}
which coincides with expression (32). Since the probabilities (32)
and (35) are equal, we can conclude that all photons, emitted by a
macroscopic ensemble of a hydrogen atoms, correspond to the
breakdown of the Einstein's equation for gravitation mass and to
quantization of the mass (17).

 Let us estimate the probability (35). If the experiment is
done by using spacecraft or satellite, then we may have
$|\phi(R'+ut)| \leq |\phi(R')|$. In this case, Eq.(35) is reduced
to Eq.(16) and can be rewritten as
 \begin{equation}
\tilde{P}_n = \biggl( \frac{V_{n,1}}{E_n - E_1} \biggl)^2
 \frac{\phi^2(R')}{c^4}  \simeq  0.49 \times 10^{-18}
 \biggl( \frac{V_{n,1}}{E_n-E_1} \biggl)^2 ,
\end{equation}
where, in Eq.(36), we use the following  numerical values of the
Earth mass, $M \simeq 6 \times 10^{24} kg$, and its radius, $R_0
\simeq 6.36 \times 10^6 m$. It is important that, although the
probabilities (36) are small, the number of photons, $N$, emitted
by macroscopic ensemble of the atoms, can be large since the
factor $V^2_{n,1}/(E_n-E_1)^2$ is of the order of unity. For
instance, for 1000 moles of hydrogen atoms, $N$ is estimated as
\begin{equation}
N_{n,1} = 2.95 \times 10^{8} \biggl(
\frac{V_{n,1}}{E_n-E_1} \biggl)^2 , \ N_{2,1} = 0.9
\times 10^8 ,
\end{equation}
which can be experimentally detected, where $N_{n,1}$ stands for a
number of photons, emitted with energy
$\hbar \omega_{n,1} = E_n -E_1$.

\section{Oscillations of the gravitational mass expectation value}

To summarize, we have demonstrated that passive gravitational mass
of a composite quantum body is not equivalent to its energy due to
quantum fluctuations and discussed a realistic indirect
experimental method to detect this difference. We have also shown
that the corresponding expectation values are equivalent to each
other for stationary quantum states. In this context, we need to
make the following comment. First of all, we stress that, for
superpositions of stationary quantum states, the expectation
values of passive gravitational mass can be oscillatory functions
of time even in case, where the expectation value of energy is
constant. For instance, as follows from Eq.(9), for electron wave
function,
\begin{equation}
\Psi_{1,2}(r,t) = \frac{1}{\sqrt{2}} \bigl[ \Psi_1(r) \exp(-iE_1t)
+ \Psi_2(r) \exp(-iE_2t) \bigl],
\end{equation}
which is characterized by the time independent expectation value of
energy, $<E> = (E_1+E_2)/2$, the expectation value of passive
gravitational mass is the following oscillatory function:
\begin{equation}
<\hat m^g_e> = m_e + \frac{E_1+E_2}{2 c^2} + \frac{V_{1,2}}{c^2}
\cos \biggl[ \frac{(E_1-E_2)t}{\hbar} \biggl] .
\end{equation}
Note that the oscillations of passive gravitation mass (39)
directly demonstrate inequivalence of passive gravitational mass
and energy at a macroscopic level. It is important that these
oscillations are strong (of the order of $\alpha^2 m_e$) and of a
pure quantum origin without classical analogs. We also pay
attention that our preliminary calculations [14] show that an
active gravitational mass can exhibit similar time dependent
oscillations of its expectation value in superpositions of
stationary quantum mechanical states. We hope that these strong
oscillations of passive and active gravitational masses are
experimentally measured, despite the fact that the quantum state
(38) decays with time.

If we average the oscillations (39) over time, we obtain the
modified equivalence principle between the averaged over time
expectation value of passive gravitational mass and the
expectation value of energy in the following form:
\begin{equation}
<< \hat m^g_e >>_t = m_e + (E_1+E_2)/2c^2 = <E>/c^2.
\end{equation}
We pay attention that physical meaning of averaging procedure in
Eq.(40) is completely different from that in classical time
averaging procedure (1),(7) and does not have the corresponding
classical analog.

\section{Summary}

In conclusion, we note that we have suggested to measure a number
of photons, emitted by excited electrons in a macroscopic ensemble
of hydrogen atoms, provided that the atoms are supported and moved
in the Earth gravitational field with constant velocity, $u \ll
\alpha c$, by spacecraft or satellite. It is important that the
physical origin of the proposed effects are due to different
couplings of kinetic and potential energies with an external
gravitational field. This proposal is very different from that,
discussed in Refs.[13,15,16], where small corrections to electron
energy levels are calculated for a free falling hydrogen atom
[15,16] or for a hydrogen atom, supported in a gravitational field
[13]. Account of the above mentioned small corrections to energy
levels cannot change significantly the number of electrons,
excited due to spacecraft or satellite motion in an inhomogeneous
gravitational field (2), calculated in the paper. We also note
that, in accordance with Ref.[17], self-forces and self-torques in
gravitational field are not important in our case, since we
consider neutral atoms.

\section*{Acknowlegements}

We are thankful to N.N. Bagmet, V.A. Belinski, Steven Carlip,
Li-Zhi Fang, Douglas Singleton, and V.E. Zakharov for very useful
discussions. This work was supported by the NSF under Grant
DMR-1104512.

\end{document}